\documentclass[twocolumn,aps,prl,amsmath,amssymb]{revtex4}

\usepackage{graphicx}

\begin{document}

\title{Few-cycle pulses from a graphene mode-locked all-fiber laser}

\author{D. Purdie$^1$, D. Popa$^1$, V. J. Wittwer$^1$, Z. Jiang$^1$, G. Bonacchini$^1$, F. Torrisi$^1$, S. Milana$^1$, E. Lidorikis$^2$, A.C. Ferrari$^1$}
\affiliation{$^1$Cambridge Graphene Centre, University of Cambridge, Cambridge CB3 0FA, UK}
\affiliation{$^2$Department of Materials Science and Engineering, University of Ioannina, Ioannina, Greece}

\begin{abstract}
We combine a graphene mode-locked oscillator with an external compressor and achieve$\sim$29fs pulses with$\sim$52mW average power. This is a simple, low-cost, and robust setup, entirely fiber based, with no free-space optics, for applications requiring high temporal resolution.
\end{abstract}

\maketitle

Ultrafast light pulses in the femtosecond range are needed for advanced photonics applications. E.g. in pump-probe spectroscopy, photophysical and photochemical relaxation processes are monitored by exciting a sample with an ultrashort light pulse. The maximum temporal resolution is determined by the duration, $\Delta\tau$, of the pulse. This is usually defined as the full width at half maximum (FWHM) of its intensity profile in the time domain, $I(t)$\cite{Fermann_ul}. Alternatively $\Delta\tau$ may be defined by the number of oscillation periods of the electric field carrier wave (optical cycles) within the pulse\cite{Kartner_fc} $N=\frac{\Delta\tau}{T_{0}}=\nu_{0}\Delta\tau$, where $T_{0}$ is the optical cycle of frequency $\nu_{0}$. The ultimate pulse duration is set by a single cycle of light, i.e. $T_{0}$, given by\cite{Kartner_fc} $\frac{\lambda}{c}$, where $\lambda$ is the wavelength and $c$ is the speed of light. Finally, the uncertainty relation $\Delta\nu\Delta\tau\simeq\frac{1}{\pi}$ provides a measure of the minimum frequency bandwidth $\Delta\nu$ required for an ultrashort pulse formation\cite{Kartner_fc}, i.e. the broader the bandwidth, the shorter the supported pulse. In the visible and near infrared (NIR), $T_{0}$ lies, e.g, between 2fs at $\lambda\sim$600nm and 5fs at $\lambda\sim$1.5$\mu$m, which set the ultimate speed limit for devices operating in this wavelength range. Achieving shorter pulses therefore requires moving to shorter wavelengths.

Pulses as short as 2-cycles can be generated directly from laser cavities using passive mode-locking\cite{Kartner_fc,Fermann_ul,Keller03n}. Ti:Saphire lasers have become established tools for few-cycle generation\cite{Kartner_fc}, with the shortest pulses produced to date having $\Delta\tau\sim$5fs\cite{Ell01ol} at a centre wavelength, $\lambda_{0}\sim$800nm, corresponding to less than 2-cycles, with spectral width $\Delta\lambda\sim$600nm\cite{Ell01ol}. Ti:Saphire lasers able to generate few-cycle durations are typically optimized to make use of the maximum $\Delta\lambda$ gain available\cite{Kartner_fc}, consequently they have no wavelength tunability\cite{Kartner_fc}. Tunable Ti:Saphire operate with a much longer pulse duration, e.g. $\Delta\tau\sim$150fs in a typical$\sim$680-1080nm commercially available spectral range\cite{Coherent}. Tunable few-cycle pulses can be achieved by exploiting nonlinear optical effects in optical parametric amplifiers (OPAs). These can be described by expressing the polarization ($P$) as a power series in the applied optical field ($E$)\cite{Boyd_no,Sutherland_hno}: $P=\epsilon_0[\chi^{(1)}E+\chi^{(2)}E^2+\chi^{(3)}E^3+...]$, where $\epsilon_0$ is the free space permittivity, $\chi^{(1)}$ is the linear and $\chi^{(2)}$ and $\chi^{(3)}$ are the second- and third-order nonlinear susceptibilities. OPAs are optical amplifiers based on the $\chi^{(2)}$ nonlinearity of a crystal\cite{Boyd_no,Sutherland_hno,Cerullo03rsi}, in a process, called parametric\cite{Boyd_no,Sutherland_hno}, where there is no net transfer of energy and momentum between $E$ and the crystal\cite{Boyd_no,Sutherland_hno}. This can be visualized, by considering energy transfer from a pump pulse of frequency $\omega_p$ to two pulses of lower frequencies $\omega_s$ and $\omega_i$, called signal and idler\cite{Boyd_no,Sutherland_hno}, with the requirement $\omega_p=\omega_s+\omega_i$\cite{Boyd_no,Sutherland_hno}. Under this condition, OPAs can transfer energy form a narrow, fixed $\Delta\lambda$, pump pulse, to a broad, variable $\Delta\lambda$, signal pulse, e.g. from$\sim$600 to over 3500nm, with pulses as short as sub-3-cycles\cite{Brida10jo,Kartner_fc}, their duration being ultimately limited by uncompensated dispersive and nonlinear effects\cite{Brida10jo,Kartner_fc}. However, both Ti:Saphire oscillators and OPAs optimized to produce few-cycle pulses are complex and expensive setups, relying on bulk optics\cite{Kartner_fc,Brida10jo}. This has driven a research effort to find novel approaches, not only capable of producing short pulses, but also cheap, simple, broadband, and inexpensive, which would make few-cycle pulses more accessible to the wider scientific community.

Compared to their solid-state counterparts, fiber lasers are attractive platforms for short pulse generation due to their simple and compact designs\cite{Fermann13np}, efficient heat dissipation\cite{Dausinger04tap}, and alignment-free operation\cite{Dausinger04tap,Kartner_fc,Fermann_ul}. These characteristics, combined with advances in glass technology\cite{Digonnet,Russel03sc} and nonlinear optics\cite{Dudley09np}, resulted in systems working from the visible to the mid-infrared (MIR)\cite{Digonnet}. In fiber oscillators, ultrashort pulses can be obtained by passive mode-locking\cite{Fermann13np}. This typically requires the aid of a non-linear component called a saturable absorber (SA)\cite{Fermann_ul, Keller03n, Fermann13np}. Graphene\cite{Hasan09am,Sun10an} and carbon nanotubes (CNTs)\cite{Scardaci08am,Wang08nn,Hasan09am} have emerged as promising SAs for ultrafast lasers\cite{Going12pe,Wang08nn,Woodward14ptl,Zhang13oe,Castellani13lpl,Popa12apl,Sun10an,Sun10nr,Popa11apl}. In CNTs, broadband operation is achieved by using a distribution of tube diameters\cite{Wang08nn,Going12pe}, while this is an intrinsic property of graphene\cite{Bonaccorso10np}. This, along with the ultrafast recovery time\cite{Brida13nc}, low saturation fluence\cite{Sun10an,Popa10apl}, and ease of fabrication\cite{Ferrari14ns} and integration\cite{Bonaccorso12mt}, makes graphene an excellent broadband SA\cite{Bonaccorso10np}. Consequently, mode-locked lasers using graphene SAs (GSAs) have been demonstrated from$\sim$800nm\cite{Baek12ape} to $\sim$970nm\cite{Zaugg13oe}, $\sim$1$\mu$m\cite{Mary13oe}, $\sim$1.5$\mu$m\cite{Popa10apl} and $\sim$2$\mu$m\cite{Zhang12oe} up to $\sim$2,4$\mu$m\cite{Cizmeciyan13ol}. Therefore, experimental setups that combine the unique optical properties and simple fabrication of graphene, with fiber lasers, are attractive prospects for few-cycle generation.

Here we achieve$\sim$29fs pulses, corresponding to less than 6-cycles at 1550nm, using a graphene mode-locked fibre oscillator in conjunction with an external compressor based on an erbium doped fiber amplifier (EDFA) and a length of Single Mode Fibre (SMF). The EDF length within the amplifier acts as the gain and non-linear medium, used to simultaneously amplify and spectrally broaden the pulse, while the SMF, placed at the output of the EDFA, acts as the dispersive delay line. Our design uses only standard telecommunication equipment and an intrinsic broadband GSA, attractive when cost, ease of fabrication, and operational stability are required. In addition, this enables a compact and portable format.

For fiber lasers, a typical approach for ultrashort pulse generation is soliton mode-locking\cite{Kartner_fc,Fermann_ul}. In this regime, the fiber dispersive and non-linear effects can cancel each other\cite{Kartner_fc,Fermann_ul}, allowing a stable pulse envelope to propagate\cite{Kartner_fc,Fermann_ul}. During propagation, the soliton periodically encounters perturbations occurring in the fiber laser of cavity length $L_{c}$\cite{Agrawal_anfo}. The shortest solitons stably supported, typically have $Z_{0}<\frac{L_{c}}{2}$\cite{Tamura93ol}, where $Z_{0}\simeq\frac{\Delta\tau^{2}}{2\beta^{(2)}}$ is the soliton period with $\beta^{(2)}$ the second-order dispersion. For pulses with $\Delta\tau<$100fs, one needs $L_{c}<\frac{\Delta\tau^{2}}{\beta^{(2)}}\simeq$50cm for fiber lasers operating at 1.5$\mu$m, where $\beta^{(2)}\simeq$-25$\frac{fs^2}{mm}$. For such short $L_{c}$, it becomes challenging to compensate dispersive effects in all-fiber formats\cite{Agrawal_anfo}. E.g., to achieve pulses$\sim$10-cycles, i.e.$\sim$30fs at 1$\mu$m\cite{Buckley06ol,Zhou08oe,Ma10ol}, the shortest reported to date\cite{Zhou08oe,Ma10ol}, bulk optics are typically employed\cite{Buckley06ol,Zhou08oe,Ma10ol} in order to compensate for dispersive effects, eliminating the advantage of alignment-free operation which makes fiber lasers attractive. A strategy to overcome these limitations in all-fiber laser formats is to use a dispersion-managed design\cite{Nelson97apb,Tamura93ol}, where alternating segments of positive (or normal) and negative (or anomalous) dispersion fibers lead to periodic broadening and compression of the intracavity pulses\cite{Nelson97apb,Tamura93ol}. Compared to soliton mode-locking, the average $\Delta\tau$ can increase by an order of magnitude or more, which significantly reduces the intracavity $P_{peak}$, thus being less susceptible to nonlinear optical effects\cite{Nelson97apb,Tamura93ol}. However, in all-fiber oscillators, it is difficult to achieve the minimum $\Delta\tau$, as set by $T_{0}$. Dispersive effects increase with $\Delta\lambda$\cite{Agrawal_anfo} and it is necessary to provide compensation for orders higher than $\beta^{(2)}$, e.g. third-order dispersion $\beta^{(3)}$ or even higher\cite{Agrawal_anfo}. In addition, the limited $\Delta\lambda$ of gain fibers, typically$\sim$50nm\cite{Digonnet_redfla}, can further affect the minimum achievable $\Delta\tau$. Thus, of particular interest are novel setups, still capable of few-cycle generation, but with simpler, more accessible designs, and ampler operation wavelength, maintaining the all-fibre design.

A different route for few-cycle generation is to externally compress the pulse\cite{Kartner_fc,Agrawal_anfo}. In this approach, the pulse is first passed through a non-linear medium, such as a SMF length\cite{Agrawal_anfo}, where it experiences self-induced phase modulation (SPM) (non-linear phase delay)\cite{Kartner_fc,Agrawal_anfo}, caused by its own intensity $I(t)$, via the Kerr effect\cite{Agrawal_anfo}, i.e. $I(t)$-dependent change in the refractive index $\Delta n = n_{2}I(t)$, where $n_{2}$ is the medium nonlinear index coefficient\cite{Agrawal_anfo}. After a propagation length $L$, the pulse accumulates a time dependent phase-shift $\Delta\phi(t)=\frac{n_{2}I(t)L\nu_{0}}{c}$\cite{Agrawal_anfo}, resulting in the generation of new frequency components $\Delta\nu(t)=\frac{1}{2\pi} \frac{d\Delta\phi(t)}{dt}$\cite{Kartner_fc,Agrawal_anfo}, distributed temporally across the pulse, i.e. the pulse becomes chirped. In order to achieve shorter $\Delta\tau$, as supported by the now broadened $\Delta\nu$, the chirped pulse is then passed through a dispersive delay line\cite{Agrawal_anfo}, which re-phases the new frequency components, $\Delta\nu(t)$, generated by the SPM\cite{Kartner_fc,Agrawal_anfo}. Ideally, the dispersive delay line would introduce the inverse of the chirp added by SPM, resulting in the compression of the pulse to its minimum duration. Typically the dispersive delay line, formed by a component with negative dispersion, compensates for the linear chirp, around the central part of the pulse, where most energy is concentrated\cite{Kartner_fc,Agrawal_anfo}. When compared to other techniques, such as OPAs, external pulse compression offers a more flexible design\cite{Kartner_fc}, with the possibility of taking advantage of the ultra-broadband spectrum which can be generated by the phase modulation processes\cite{Kartner_fc,Agrawal_anfo} in all-fiber formats\cite{Agrawal_anfo}.
\begin{figure}[tb]
\centerline{\includegraphics[width=8.6cm]{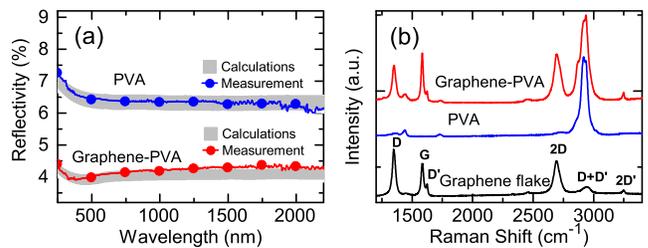}}
\caption{(a) Calculated and measured reflectivity of PVA reference and graphene-PVA composite. The grey areas represent the combined results of$\sim$2000 random graphene orientations within the composite. (b) 514.5nm Raman spectra of a graphene flake, PVA reference and graphene-PVA composite.}
\label{fig:transmittance}
\end{figure}

We use a graphene-PVA composite as our GSA, fabricated via solution processing. Graphite flakes are exfoliated from bulk graphite by mild ultrasonication with sodium deoxycholate (SDC) surfactant\cite{Sun10an,Bonaccorso12mt}. A dispersion enriched with single layer (SLG) and few layer graphene is then mixed with an aqueous solution of polyvinyl alcohol (PVA). After water evaporation, a graphene-PVA composite is obtained\cite{Sun10an}. The reflectivity measurements of the graphene-PVA composite and the PVA reference are presented in Fig.\ref{fig:transmittance}(a). We estimate the number of graphene layers (N) present in the composite from the reflectivity measurements, by using the transfer matrix formalism\cite{Born}. We calculate, as a function of N, the reflectivity of a PVA-graphene composite with overall thickness corresponding to the experimental one, and with graphene layers randomly distributed within the matrix. The PVA refractive index n($\lambda$) is adjusted to reproduce the experimental reflectivity measured on the pure polymer film in the VIS-IR range, n$\sim$1.44\cite{Cheremisinoff}. In order to avoid coherent multiple reflections due to a specific arrangement of the graphene layers in the composite, we perform a statistical sampling by repeating each calculation for many ($\sim$2000) random graphene orientations within the film. By comparing our calculations with the experimental reflectivity, Fig.\ref{fig:transmittance}(a), we estimate that a 4\% overall reflectivity translates to N$\sim$30-35. Fig.\ref{fig:transmittance}(b) plots the Raman spectra of a graphene flake deposited on Si/SiO$_{2}$, the PVA reference and the graphene-PVA composite. Besides the G and 2D peaks, the Raman spectrum of the flake has significant D and D' intensities\cite{Ferrari00prb,Ferrari06prl}. We assign the D and D' peaks to the edges of the submicrometer flakes\cite{Torrisi12an}, rather than a large amount of disorder within the flakes\cite{Casiraghi09nl}. This is further supported by analyzing the G peak dispersion, Disp(G). In disordered carbons the G peak position, Pos(G), increases with decreasing excitation wavelength, from IR to UV\cite{Ferrari00prb}. Thus, Disp(G)=$\Delta Pos(G)/\Delta \lambda_{0}$, increases with disorder\cite{Ferrari01prb,Ferrari00prb}. FWHM(G) always increases with disorder\cite{Cancado11nl}. Hence, combining the intensity ratio of the D and G peaks, I(D)/I(G), with FWHM(G) and Disp(G) allows us to discriminate between edges, and disorder in the bulk of the samples. In the latter case, a higher I(D)/I(G) would correspond to higher FWHM(G) and Disp(G). By analyzing 30 flakes, we find that the distribution of Disp(G), I(D)/I(G) and FWHM(G) are not correlated, indicating that the D peak is mostly due to edges. Also, Disp(G) is nearly zero for all samples (compared to$\geq$0.1cm$^{-1}$/nm expected for disordered carbons\cite{Ferrari01prb}). Although 2D is broader than in pristine graphene, it is still a single Lorentzian. This implies that even if the flakes are multilayers, they are electronically decoupled and, to a first approximation, behave as a collection of single layers\cite{Latil07prb}. The spectrum of the graphene-PVA composite (Fig.\ref{fig:transmittance}(b)) can be seen as a superposition of that of the flake and PVA. Thus, PVA does not affect the structure of the embedded flakes.
\begin{figure}[tb]
\centerline{\includegraphics[width=8.6cm]{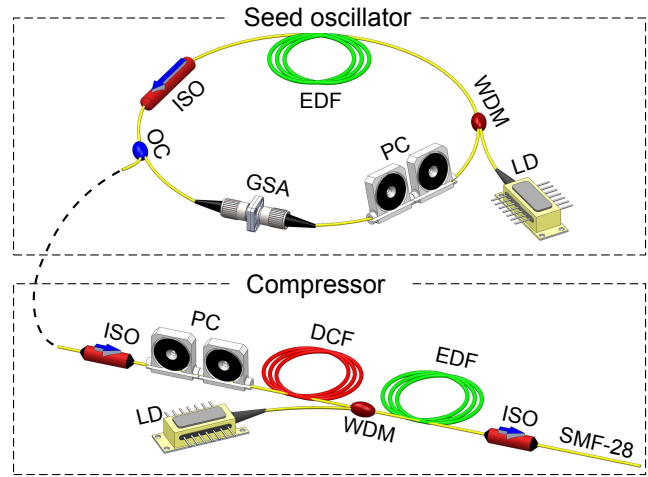}}
\caption{Laser schematic. The seed oscillator is mode-locked using a GSA to produce 263fs seed pulses. These are fed through the compressor, reducing their duration down to 29fs. LD: Laser diode; WDM: Wavelength division multiplexer; EDF: Erbium doped fiber; ISO: Isolator; OC: Optical coupler; PC: Polarization controller; DCF: Dispersion compensation fiber; SMF: Single mode fiber.}
\label{fig:experimental_setup}
\end{figure}

\begin{figure}[tb]
\centerline{\includegraphics[width=8.6cm]{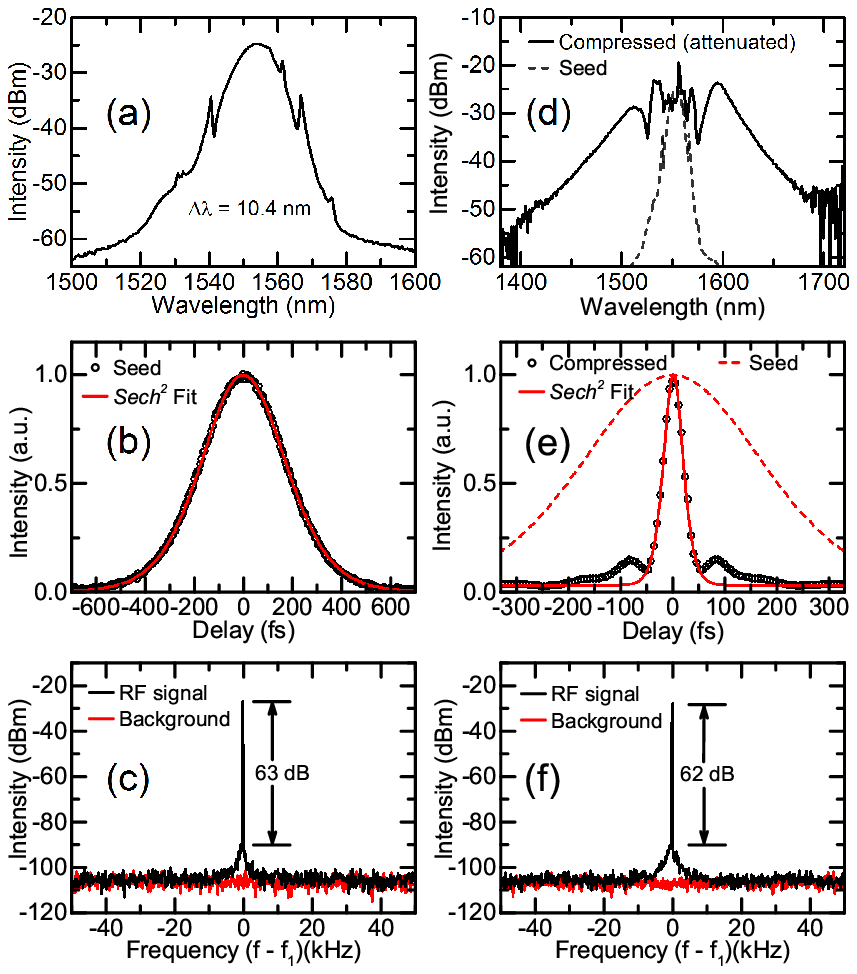}}
\caption{Seed pulse characterisation. (a) Optical spectra, with FWHM=$\Delta \lambda_{s}$=10.4nm. (b) Intensity autocorrelation trace fitted with a \textit{sech$^2$} profile, giving FWHM=$\Delta\tau_{s}$=263fs after deconvolution. (c) RF spectrum showing the first harmonic, measured around $f_{1}=$18.67MHz with a 100kHz span and 30Hz resolution. (d) Optical spectra. (e) Intensity autocorrelation trace, fitted with a \textit{sech$^2$} profile. Both seed and compressed traces are normalized to 1. (f) RF spectrum, after attenuation to$\sim$1.5mW to avoid damage to our photodiode, showing the first harmonic at 18.67MHz, with a 100kHz span and 30Hz resolution}
\label{fig:results}
\end{figure}

For our seed oscillator, we design a dispersion-managed soliton\cite{Popa10apl}, able to generate shorter $\Delta\tau$ with broader $\Delta\lambda$ than soliton lasers\cite{Nelson97apb}, schematically presented in Fig.\ref{fig:experimental_setup}(a). The total cavity length is $L_c$$\sim$11m. We use a $L_1$$\sim$3.7m EDF, with $\beta^{(2)}_1\sim23\frac{ps^2}{km}$ as gain medium. The rest of the cavity is formed from two lengths of standard SMF: $L_2\sim$6.9m of SMF-28 with $\beta^{(2)}_2\sim-22\frac{ps^2}{km}$, and $L_3\sim$40cm of Flexcore-1060 with $\beta^{(2)}_3\sim-7\frac{ps^2}{km}$. This gives a net intracavity second-order dispersion $L_1\beta^{(2)}_1$+$L_2\beta^{(2)}_2$+$L_3\beta^{(2)}_3\sim$ -0.07ps$^2$, typical of dispersion-managed soliton lasers\cite{Popa10apl,Popa12apl}. The output of the cavity is the 30\% port of a $\frac{70}{30}$ coupler. The GSA is placed after the coupler. Fundamental mode-locking is achieved with an output power $P_{out}$=0.68mW and repetition rate $f_{1}$=18.67MHz. The seed pulse optical spectrum is shown in Fig.\ref{fig:results}(a), with $\Delta\lambda_{s}=$10.4nm. The corresponding intensity autocorrelation trace is shown in Fig.\ref{fig:results}(b), with $\Delta\tau_{s}$=263fs, as determined by fitting a \textit{sech$^2$} profile to the pulse, as expected for soliton-like mode-locking\cite{Agrawal_anfo}. This gives a time bandwidth product (TBP) $\Delta\nu_{s}\Delta\tau_{s}$=0.34, close to the expected transform-limit 0.315\cite{Keller04ultra}.

The design of the compressor is based on an EDFA, as shown in Fig.\ref{fig:experimental_setup}(b). To protect against back reflections\cite{Becker99edfa}, two isolators are placed at both the input and output, consisting of$\sim$40 and $\sim$50cm of SMF-28. To reduce $P_{peak}$ and avoid temporal pulse-shape distortions\cite{Fermann_ul} or damage to the compressor components\cite{Fermann_ul}, we stretch the pulse using$\sim$3.6m of dispersion compensated fiber (DCF) with $\beta^{(2)}\sim60\frac{ps^2}{km}$, after the input isolator. After the DCF, a WDM consisting of$\sim$220cm Flexcore-1060, is used to forward pump$\sim$3m of EDF with $\beta^{(2)}\sim17\frac{ps^2}{km}$, used as the gain medium. Finally, a length of SMF-28, placed after EDF, forms the dispersive delay line, for which an optimized length$\sim$125cm (including the$\sim$50cm output isolator) is used, as determined by monitoring the autocorrelation trace at the compressor output until a minimum in the pulse duration is achieved. A polarization controller (PC) is used to match the polarization state of the incident seed pulses (compression optimization).

The EDFA within the compressor is operated with a pump power $P_{p}\sim$350mW. Fig.\ref{fig:results}(a) plots the optical spectrum recorded at the output of the compressor. $\Delta\lambda_{c}>$100nm indicates spectral broadening has occurred. The corresponding intensity autocorrelation trace is shown in Fig.\ref{fig:results}(b), with a \textit{sech$^2$} profile of $\Delta\tau_{c}\sim$29fs, i.e. $\sim$6 optical cycles. This gives a compression factor $\frac{\Delta\tau_{s}}{\Delta\tau_{c}}=9$. The radio frequency (RF) spectrum of the pulse train, measured with a photodetector connected to an RF spectrum analyzer, is reported in Figs.\ref{fig:results}(c,d), before and after compression, respectively. Relative to the seed pulses, the compressed pulses have no degradation in signal-to-noise ratio (SNR), indicating pulse stability is maintained during compression\cite{Linde86apb}. After compression, $P_{out}$=52mW, corresponding to a pulse energy $E_{c}$=2.8nJ and $P_{peak}$=85kW. An increased pedestal is also observed at the base of the pulse, which can be attributed to uncompressed non-linear chirp added to the pulse through SPM\cite{Agrawal_anfo}. Compensation of higher order chirp terms would enable further reduction in pulse duration\cite{Kartner_fc,Agrawal_anfo}, as well as minimization of the pedestal structure at the base of the pulse\cite{Agrawal_anfo}. This can be achieved through control of the higher order dispersion terms within the dispersive delay line. E.g. to control $\beta^{(3)}$\cite{Agrawal_anfo}, prism pairs or chirped mirrors\cite{Kartner_fc}, or lengths of photonic crystal fibers (PCFs)\cite{Russel03sc}, could be used. The amount of spectral broadening achievable with our setup, therefore the minimum supported pulse duration, is further limited by the maximum $P_{p}$ of the laser diode used to pump the EDFA. Use of a higher power diode, or a double pumped configuration, should enable the generation of higher bandwidth pulses, as well as enabling increased output power. Finally, wavelength-engineered dispersion management, e.g. using PCFs\cite{Russel03sc}, coupled with the broadband nature of graphene\cite{Bonaccorso10np}, could, in principle, extend our approach to enable short pulses at other wavelengths.

Compared to Ti:Sapphire and OPA designs, capable of producing $\sim$2-cycle pulses\cite{Brida10jo,Kartner_fc}, our all-fiber design, exploiting a GSA, represents a simple to assemble all-fiber components, needing no critical alignment. A CNT-based fiber setup capable of $\sim$14fs pulses at 1550nm was reported in Ref.\onlinecite{Kieu10ptl}. Despite the shorter pulses, Ref.\onlinecite{Kieu10ptl} used bulk optics (a prism pair)\cite{Kieu10ptl}, unlike the alignment-free all-fiber format of our laser. Also, a CNT-based SA limits operation wavelength, compared to our GSA, that can operate at any wavelength.

In conclusion, we reported a graphene-mode locked laser generating 29fs pulses with an average output power$\sim$52mW and pulse energy 2.8nJ, corresponding to a peak power$\sim$85kW, making it attractive for applications such as optical frequency comb generation and high resolution laser spectroscopy.

We acknowledge funding from EU Graphene Flagship (no. 604391), ERC Grant Hetero2D, EPSRC Grants EP/K01711X/1, EP/K017144/1, EP/M507799/1, EP/L016087/1, a Royal Society Wolfson Research Merit Award and Emmanuel College, Cambridge.

\end{document}